\documentclass[journal]{IEEEtran}

\pdfoutput 1

\usepackage[pdftex]{graphicx}
\graphicspath{{./Figures/}}
\usepackage{epsfig}
\usepackage{graphicx}
\usepackage{algorithmic}
\usepackage{array}
\ifCLASSOPTIONcompsoc
  \usepackage[caption=false,font=normalsize,labelfont=sf,textfont=sf]{subfig}
\else
  \usepackage[caption=false,font=footnotesize]{subfig}
\fi
\usepackage{dblfloatfix}

\usepackage{amsmath, bbm} 
\usepackage{amssymb}  
\usepackage{arydshln}
\usepackage{todonotes}
\usepackage{url}
\usepackage{tabulary}

\usepackage{amsfonts}
\usepackage{gensymb}
\usepackage{todonotes}

\usepackage{tikz}
\usepackage{tikz-qtree}
\usetikzlibrary{trees} 

\newcommand{\R}{\mathbb{R}}
\newcommand{\N}{\mathcal{N}}

\newcommand{\y}{\mathbf{y}}
\newcommand{\Pb}{\mathbf{P}}
\newcommand{\Q}{\mathbf{Q}}

\newcommand{\s}{\pmb{s}}
\newcommand{\Scap}{\pmb{S}}

\renewcommand{\Re}{\operatorname{\textbf{Re}}}
\renewcommand{\Im}{\operatorname{\textbf{Im}}}

\begin{document}
%
\title{Linear Single- and Three-Phase Voltage Forecasting and Bayesian State Estimation with Limited Sensing}

%
%
%

\author{Roel~Dobbe, Werner~van~Westering, Stephan~Liu, Daniel~Arnold, Duncan~Callaway~and~Claire~Tomlin~-~\IEEEmembership{Fellow,~IEEE}
\thanks{R. Dobbe is with the AI Now Institute at New York University, e-mail: roel@ainowinstitute.org. 
W. van Westering is with Alliander, a distribution utility in The Netherlands.
D. Arnold is with the Grid Integration Group at Lawrence Berkeley National Laboratory.
D. Callaway is with the Energy \& Resources Group at UC Berkeley. S. Liu and C. Tomlin are with the Department
of Electrical Engineering \& Computer Sciences, UC Berkeley.
Manuscript received June 2018.}}

\maketitle

\begin{abstract}
Implementing state estimation in low and medium voltage power distribution is still challenging given the scale of many networks and the reliance of traditional methods on a large number of measurements.
This paper proposes a method to improve voltage predictions in real-time by leveraging a limited set of real-time measurements.
The method relies on Bayesian estimation formulated as a linear least squares estimation problem, which resembles the classical weighted least-squares (WLS) approach for scenarios where full network observability is not available.
We build on recently developed linear approximations for unbalanced three-phase power flow to construct voltage predictions as a linear mapping of load predictions constructed with Gaussian processes.
The estimation step to update the voltage forecasts in real-time is a linear computation allowing fast high-resolution state estimate updates. 
The uncertainty in forecasts can be determined a priori and smoothed a posteriori, making the method useful for both planning, operation and post-hoc analysis.
The method outperforms conventional WLS and is applied to different test feeders and validated on a real test feeder with the utility Alliander in The Netherlands.
\end{abstract}

\begin{IEEEkeywords}
voltage forecasting, state estimation, Gaussian processes, limited sensing, linear least squares estimation
\end{IEEEkeywords}

%
\IEEEpeerreviewmaketitle

%

\section{Introduction}

The operation of electric distribution networks is faced with new challenges due to the rapid adoption of distributed generation (DG) and the electrification of our society, such as in driving and heating.
The inherent intermittency of renewable generation combined with the diversification of demand make power flow  more variable and harder to predict, leading to new protection issues, such as unintended islanding or tripping~\cite{kaur_effects_2006}, and economic burden due to accelerated wear~\cite{seltenrich_new_2013}. 
In  addition, the increasing number of connected control devices on networks expose new vulnerabilities for cyber-attacks requiring new mitigation strategies~\cite{zhao_short-term_2017}.
To understand and mitigate these risks, many Distribution System Operators (DSOs) are building a stronger information layer on top of their physical infrastructure to exploit recent advances in sensing, communication and algorithms.
Firstly, DSOs gather historical data from SCADA and AMI systems to enable \emph{forecasting} of demand, flow and voltage variables. 
Unfortunately, the increasing variability of power yields probability distributions with long tails, which cause forecasting methods to do poorly in situations when observability is most needed; when extreme and potentially dangerous events happen.
Secondly, DSOs are increasingly applying traditional \emph{state estimation} methods to their distribution networks. 
Power system state estimation (SE) is the process of leveraging measurement from a subset of states in an electric network to estimate states that are not measured in real-time.
In transmission systems, the need for system reliability has long motivated the development of state estimation methods \cite{abur_power_2004, giannakis_monitoring_2013}.
In the traditional setting, a state estimator relies on an \emph{overdetermined} formulation for the estimation problem to be well-conditioned such that the unknown/unmeasured variables are \emph{observable}. This means that the number of available measurements must be greater than or equal to the number of unknowns (to be estimated). 
Unfortunately, ensuring observability requires DSOs to equip most buses in a network with real-time sensors and communication infrastructure, leading to steep investments that are hard to scale across all territories.
In addition, if \emph{topological} observability is achieved for traditional WLS methods during the design stage~\cite{monticelli_network_1985-1}, this is not a guarantee for \emph{numerical} observability, meaning that ill-conditioned problems may still arise during the execution in operation~\cite{korres_optimal_2015}.

In this paper, we address the context of having both limited data for forecasting and limited real-time sensing capabilities, a problem faced widely by DSOs.
In this situation, DSOs struggle to build a reliable voltage information source for both planning or real-time operation purposes.
Rather than trying to build a fully observable state estimation problem in the classical sense, one may use a statistical learning approach to use a limited set of high-resolution measurements, such as those collected by Phasor Measurement Units (PMUs)~\cite{von_meier_micro-synchrophasors_2014}, to \emph{update} a forecast initially built with historical data.
It turns out that with appropriate linear models of power flow such a Bayesian operation can be done efficiently and adequately using linear algebra operations only. The resulting \emph{linear least squares estimation} approach builds upon earlier work~\cite{dobbe_real-time_2016} and has some related previous works~\cite{schenato_bayesian_2014,weng_probabilistic_2017}.
However, we will discuss our method in the context of the classical literature on state estimation, looking in particular at other methods that combine measurements and data from different sources to construct full-network estimates. 
We explain how our method differs to current literature gaps and practical challenges.

\subsection{Previous Work}

In conventional state estimation, the measurements $z \in \R^{N_m}$ are expressed as a function of the quantities that are estimated $x \in \R^{N_n}$, by using power flow modeling:
\begin{equation}
z = h(x) \,.
\label{eq:measurementeq}
\end{equation}
The state estimation problem is then most often solved using a weighted least squares (WLS) problem:
\begin{equation}
x^* = \arg \min_{x} (z - h(x))^{\top}W(z - h(x)) \,,
\label{eq:wls}
\end{equation}
where $W$ is a matrix putting different weights on different measurements based on noise or measurement quality information~\cite{abur_power_2004}.
For the WLS problem to yield a meaningful result, the equation in \eqref{eq:measurementeq} needs to be \emph{overdetermined}, $N_m > N_n$. 
A key challenge in distribution grids is to estimate an $N_n$-dimensional state vector in scenarios where only a limited set of sensors is available, i.e. $N_m < N_n$, which does not satisfy the requirements for conventional state estimation. As a result, the standard estimation problem is \emph{underdetermined} and hence ill-posed. In practice, this means that the state vector $x$ is not \emph{observable}~\cite{monticelli_network_1985-1}.
To overcome this challenge, \emph{pseudo-measurements} are typically used to augment real-time measurements in a weighted least squares (WLS) estimation algorithm, which are often calculated using load forecasts or historical data that tend to be less accurate than real-time measurements.
Initial efforts considered augmenting an already fully observed measurement vector with extra load forecasts \cite{rousseaux_whither_1990, blood_electric_2008}.
Later efforts tried to use a more limited number of real-time measurements with forecasts from Gaussian Mixture Models~\cite{singh_distribution_2010} or Artificial Neural Networks~\cite{manitsas_distribution_2012}.
In transmission networks, the advent of phasor measurement units (PMUs) has spurred many efforts aiming to integrate high-frequency PMU measurements with lower frequency measurements from traditional sensors, yielding various \emph{hybrid} estimators that can track faster dynamics during normal operations and when contingencies occur~\cite{guo_multi-time_2015,zhao_power_2016,asprou_two-stage_2017,manousakis_hybrid_2018}.
The use of forecasting has been around to predict the evolution of \emph{dynamic states}, often as a result of load behavior, to further enhance a given static estimate~\cite{da_silva_state_1983,do_coutto_filho_forecasting-aided_2009}.

In distribution networks, the need for state estimation and the advent of PMUs are a more recent phenomenon.
Many contributions have been made to enable Distribution System State Estimation based on traditional WLS, and we refer the reader to \cite{primadianto_review_2017} for a rigorous overview.
In networks where a SCADA system is available, a limited set of high-frequency PMU time series can be integrated with pseudo measurements approximated from the last SCADA update, typically operating at slow timescales on the order of minutes. 
G\"{o}l and Abur address this challenge through combining WLS with a least absolute value step that yields a hybrid method that is robust to error~\cite{gol_hybrid_2015}.
Unfortunately, many networks still lack a SCADA infrastructure or have data collections that are deficient.
In such cases, the use of PMUs in state estimation may still be possible by resorting to load forecasting information to provide pseudo-measurements. 
Our earlier work~\cite{dobbe_real-time_2016}, and work by Schenato~\cite{schenato_bayesian_2014} and Weng~\cite{weng_probabilistic_2017} studies the use of Bayesian estimation for state estimation, using load statistics to determine \emph{prior} probabilistic forecasts $\hat x$ of state variables, which can be updated based on a limited set of real-time measurements.
These papers show the accuracy is comparable to that of conventional WLS estimators, and estimation error confidence intervals can be computed off-line, allowing for engineering trade-offs between number of sensors and estimation accuracy.

\subsection{Contributions}
This paper builds upon the contributions in~\cite{dobbe_real-time_2016} to enable state estimation for steady-state conditions by improving state forecasts with limited sensing using a Bayesian approach. 
Concretely, we build time-series forecasting methods that provide both point estimates (mean) and uncertainty (variance) for all voltages by integrating recent advances in linear models for three-phase power flow to enable our method in more complex distribution scenarios~\cite{sankur_linearized_2016,sankur_optimal_2018}. Having the two statistical moments (mean and variance) available for all voltages then allows us to formulate a Bayesian Linear Least Squares Estimation problem, through which the forecasts of all non-measured voltages can be updated based on a limited set of voltage measurements. 
The final result is a \emph{closed-form analytic state estimator} that takes as its inputs load forecasting information, a network model and real-time measurements from a limited set of sensors. 
The method works for any number of real-time sensors, and is particularly useful in many domains where there are not enough resources to install a large set of sensors.
Applied and assessed on a specific IEEE test feeder, we show that the method reduces the error of forecasts by an average 60\%, with more dramatic improvements for specific buses where forecasts are not able to perform appropriately.
Lastly,we implement the method on a real network with the DSO Alliander in The Netherlands, showcasing its use in situations where little measurement (both historical data and real-time sensing) is available.

\subsection{Notation}

We use $\| \cdot \|$ to denote the $\ell_2$-norm,  $(\cdot)^*$ to denote an optimal value, and $^{\top}$ stands for the transpose operator.
$\mathfrak N(\mu,\sigma^2)$ denotes a normal distribution with mean~$\mu$ and variance~$\sigma^2$.
Throughout this work, we use the symbol $\circ$ to represent the Hadamard Product of two matrices (or vectors) of the same dimension, also known of the element-wise product, such that:
\begin{equation*}
	C = A \circ B = B \circ A \Rightarrow C_{ij} = A_{ij} B_{ij} = B_{ij} A_{ij}
    \label{eq:hadamard}
\end{equation*}

\noindent where $i$ indicates the row and $j$ indicates the column of the vector or matrix.
\label{sec:introduction}

\section{Methodology}

\begin{figure*}[h!]
  \centering
  \epsfig{width=\textwidth,file=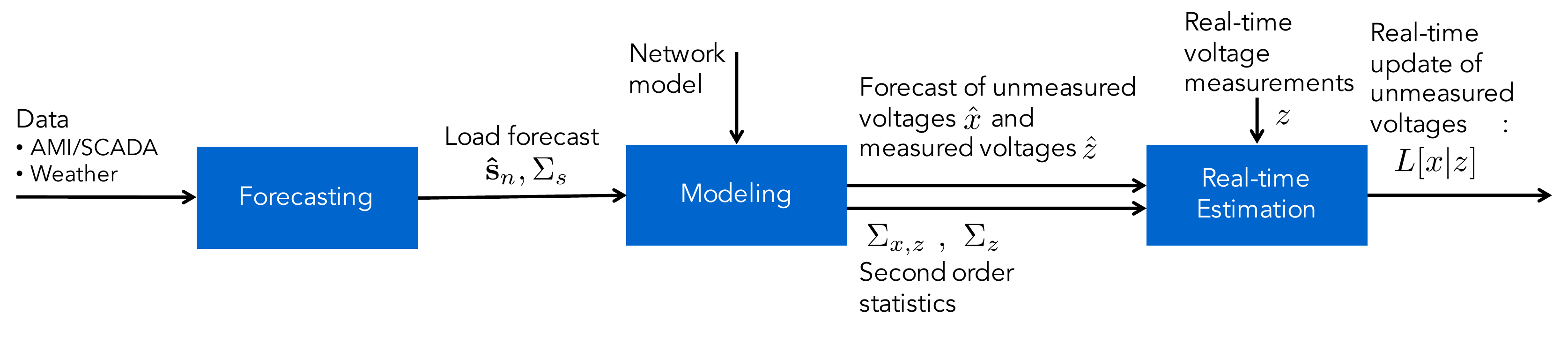}
  \caption{Overview of the forecasting and state estimation methodology.}
  \label{fig:overview}
\end{figure*}

This paper proposes a data-driven approach to do state estimation, that relies on minimum mean squares estimation (MMSE)~\cite{walrand_probability_2014}. MMSE is related to weighted least squares, but grounded in Bayesian principles and does not require an overdetermined measurement equation. 
Instead, our method relies on linear power flow models that enable us to express \emph{voltage differences} (both magnitude and angle) throughout a network as a function of nodal load and generation. 
By expressing both the measured differences and the estimated differences as a function of the load, we are able to set up a linear least squares estimation (LLSE) problem, the linear version of MMSE.
The LLSE has an analytical solution that can update the forecast of non-measured voltage differences by comparing the measured voltage differences against their forecasted values. 
As such, the method reminds of the Kalman Filter~\cite{kalman_new_1960}, which is a repeated execution of LLSE problems taking into account the potential dynamic evolution of state variables.
The approach comes with a trade-off, as the quality of the updates depends on the number of sensors and their placement in the network.

The MMSE approach enables an end-to-end pipeline from historical load and network data to voltage forecasting to updating these forecasts in real-time using a limited set of sensors. The methodology is depicted in Fig. \ref{fig:overview}.
The three main steps of forecasting, modeling and real-time estimation are developed in Sections~\ref{sec:forecasting}, \ref{sec:modeling} and \ref{sec:estimation} respectively. Before we dissect these steps we first cover the sources of uncertainty in state estimation and we introduce MMSE.

\subsection{Sources of information and uncertainty for state estimation}
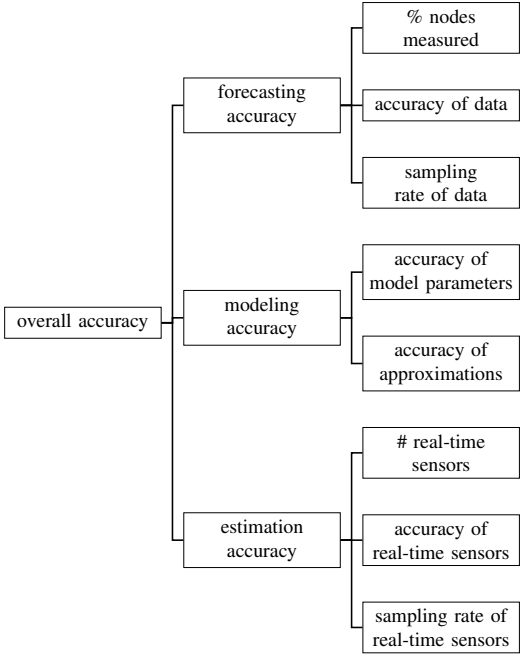
\begin{figure}
\centering
\begin{tikzpicture}[level distance=1.25in,sibling distance=.25in,scale=.75]
\tikzset{edge from parent/.style= 
            {thick, draw,
                edge from parent fork right},every tree node/.style={draw,minimum width=1in,text width=1in, align=center},grow'=right}
\Tree 
    [. {overall accuracy}
        [.{forecasting accuracy}
            [.{\% nodes measured } ]
            [.{accuracy of data } ]
            [.{sampling rate of data } ]
        ]
        [.{modeling accuracy} 
            [.{accuracy of model parameters} ]
            [.{accuracy of approximations} ]
        ]
        [.{estimation accuracy}
            [.{\# real-time sensors } ]
            [.{accuracy of real-time sensors } ]
            [.{sampling rate of real-time sensors } ]
        ] 
    ]


\end{tikzpicture}
\caption{Sources of uncertainty affecting the accuracy of the state estimator} \label{fig:sources_uncertainty}
\end{figure}

Fig.~\ref{fig:sources_uncertainty} outlines the sources of uncertainty DSOs face in constructing voltage forecasts or state estimator.
Following the proposed construction of the state estimator as depicted in Fig.~\ref{fig:overview}, the overall accuracy of the available information for state estimation depends on three sources of uncertainty: accuracy of forecasted quantities, of the modeling procedure and of quantities measured in real-time. 

The forecast may be based on a DSO's historical data, which can include SCADA data of network variables, advanced metering infrastructure (AMI) readings of household consumption (or an anonymized/aggregated version of these), data of distributed generation and storage, public weather data (temperature, humidity, solar irradiance). These data sources typically do not form a perfect representation to forecast all the necessary quantities in the network. Certain nodes may not be recorded, the recordings may be noisy or miss certain data points, and the sampling rate of the recordings may be lower than the anticipated rate for updating the state estimator.
In the modeling step, inaccuracy arises from model parameters that are outdated or identified under noisy conditions, or the bias introduced by model approximations (such as linear power flow).
Lastly, for the actual estimation, we rely on a limited number of sensors, which may be subject to measurement noise and which may have various sampling rates.

\subsection{Introduction to Minimum Mean Square Estimation}
\label{subsec:mmse}
Consider having a set of voltage phasor measurements $Z \in \mathbb R^{N_m}$ and an unobserved random variable $X \in \mathbb R^{N_n}$, representing all non-measured voltage phasors. 
We aim to determine an estimate of $X$ based on $Z$ that is close to $X$ in some sense.
Assume we are given a joint distribution of $(X,Z)$. 
We want to find an estimator $\hat X = g(Z)$ that minimizes the mean square error $E[\| X - \hat X \|^2]$. 
One can show that the minimum mean squares estimate (MMSE) of $X$ given $Z$ is equivalent to the conditional expectation $\hat X = E[X | Z]$,~\cite{walrand_probability_2014}.

We consider the case in which both the estimator and the measurements are \emph{linear} in a shared set of variables for which distributions are available, in our case in the form of load statistics. 
Let $(X,Z)$ be vectors of random variables on some probability space. 
It turns out that the estimator minimizing the mean square error is also linear in the measurements, i.e. the \emph{linear least squares estimator (LLSE)} has the form
\begin{equation}
L[X | Z ] = E[X] + \Sigma_{X,Z} \Sigma^{-1}_Z (Z - E[Z]) \,,
\label{eq:llse}
\end{equation}
where $\Sigma_{X,Z} \in \mathbb R^{N_n \times N_m}$ and $\Sigma_Z \in \mathbb R^{N_m \times N_m}$ denote the cross-covariance matrix of $X$ and $Z$ and covariance matrix of $Z$.
Interpreting \eqref{eq:llse}, $(Z - E[Z])$ represents a deviation of the actual measurement $Z$ from its expected value $E[Z]$, which is called an \emph{innovation}. This innovation triggers the Bayesian estimator $L[X|Z]$ to propose an update of the forecast $E[X]$ by a linear scaling through the covariance matrices.
Alternatively, $L[X|Z] = g(Z)$ can be interpreted as a \emph{projection} of $X$ onto the set of affine functions of $Z$.

The LLSE has four important benefits. 
Firstly, it has an analytical closed-form solution that can be used to neatly integrate real-time measurements $Z$ and forecast information (as we will see in Section \ref{sec:forecasting}).
Secondly, it is not necessary to explicitly calculate the Bayesian posterior probability density function over $X$, because $L[X|Z]$ only depends on the first two moments of $X$ and $Z$, i.e. the mean and variance.
Thirdly, it works for many distributions $(X,Z) \sim \mathfrak D$, as long as $\mathfrak D$ has well defined first and second moments~\cite{walrand_probability_2014}. 
Lastly, the number of measurements $N_m$ does not need to be larger than the number of to-be-estimated states $N_n$, which is the most significant difference with other ubiquitous estimation schemes such as weighted least squares and Gauss-Markov estimation that do not work for $N_m < N_n$. 
A challenge of any MMSE approach is understanding what information is lost in the projection that happens in~\eqref{eq:llse} through the mapping $\Sigma_{X,Z} \Sigma^{-1}_Z$.
For our state estimation method this requires revisiting the notion of network observability, typically defined for situations where $N_m > N_n$.
\label{sec:methodology}

\section{Forecasting}

We consider the design of a machine learning model to forecast the mean $\mu_s$ and covariance matrix $\Sigma_s$ of the load $\s$, which are then used to forecast the mean and covariance of the voltage magnitude and phase.
In practical contexts, DSOs may not have access to voltage or load readings from AMI in real-time, but it is possible that historical readings are used, in combination with other predictive covariates, to predict load values for a future time.

Machine learning models have been used in a variety of ways to predict load values~\cite{mirowski_demand_2014}. Two relevant examples are autoregressive moving average (ARMA) models for short term load forecasting and data-driven modeling of physical systems that utilizes regression trees to predict loads, with notable benefits to both. 
ARMA models capture trends in previous datapoints \cite{huang_short-term_2003}, but are often not practical in distribution operation, since the AMI data is mostly not available in real-time, preventing the use of recent load values.\
A regression tree model is able to cluster data based on certain characteristics, such as day of the week, temperature, and humidity \cite{behl_data-driven_2016}. 
Its interpretability makes it useful in contexts where operators need to make decisions based on a model's predictions.

In our setting, the MMSE estimator defined in Section \ref{subsec:mmse} necessitates the input of a point estimate of the load and its covariance matrix. 
This requirement motivates the use of Gaussian Processes (GPs), which offer both mean and variance information~\cite{rasmussen_gaussian_2004}.
A GP is are also flexible in that they can have continuous ARMA features as well as dicrete features as its inputs.
GPs have previously been used in similar applications for short term load forecasting to predict maximum daily loads \cite{mori_probabilistic_2005}.
Using GPs does introduce some bias, as load distributions tend to be non-Gaussian, though typically near-unimodal. In our setting, this bias is partly compensated by the LLSE. 

Let $\mathcal{N}$ denote a set of buses indexed by $n = 0,1,\dots,N-1$, where $N$ is the order (number of nodes) of the distribution feeder, and node 0 denotes the feeder head (or substation).
For each node $n \in \N$, we start with a data set of historical readings of inputs $X_n = \{ x_n[t] \in \mathcal X_n \}_{t = 1}^T$  and load values $S_n = \{ \s_n[t] \in \mathcal Y_n \}_{t = 1}^T$. 
The inputs consist of real-valued and discrete-valued features. We consider the following real-valued features at time $t$:
\begin{equation}\label{eq:feature_vec}
\left[ l_n[t-1] \cdots  l_n[t-k], \ d_n[t-1]  \cdots  d_n[t-k+1], \ \theta[t], \ \eta[t] \right],
\end{equation}
where $l_n[t]$ denotes the load value for bus $n$ at time $t$, $d_n[t]$ the difference in load between time $t-1$ and $t$, and $\theta_t$ and $\eta_t$ are the temperature and humidity at time $t$.
Note that a typical distribution feeder SCADA system often does not have access to load measurement, and hence the features $l$ and $d$ may only be available historically or in real-time for only a subset of the buses.
Hence, we also consider discrete-valued features representing date and time:
\begin{equation}\label{eq:feature_vec2}
\begin{bmatrix} DST & MOY & BD & DOW & HOD & MOH \end{bmatrix} \,,
\end{equation}
which respectively denote an indicator for daylight saving time, month of year, an indicator for business day, day of week, hour of day and minute of hour. 

We now want to train a function $f_n \ : \ \mathcal{X}_n \to \mathcal{Y}_n$ with data that best predicts $\s_n[t]$ at some time $t$ based on an input with accessible inputs $x_n[t]$.
A GP defined on an input space $\mathcal{X}_n$ can be formulated as
\begin{equation}
    f_n(x) = \pmb{\phi}_n(x_n)^{\top}\beta_n + g_n(x) \,,
\end{equation}
where $g_n(x)$ is a zero-mean GP represented as $\mathcal{GP}(0,k_n(x_n,x_n))$, with kernel $k_n(x_n,x_n)$ modeling the covariance across the input space~$\mathcal{X}_n$.
$\pmb{\phi}_n(x_n)^{\top}\beta_n$ determines the translation of the GP from the origin, with $\pmb{\phi}_n(x_n)$ a feature basis for the output given the input vector $x_n$, $\beta_n$ are learned coefficients or weights for the basis features \cite[Section 2.2]{rasmussen_gaussian_2004}. 
Given this framework, we can model the distribution of an output at a certain input $x_n^*$: 
\begin{equation}\label{eq:gp_dist}  f(x_n^*) \ | \ x_n^*, X_n,S_n \sim \mathfrak{N}( \pmb{\phi}_n(x_n^*)^{\top}\beta_n + g_n(x_n^*) , \sigma^2) \,,
\end{equation}
The primary assumption under GPs is that it models a collection of random variables, any finite number of which have a joint Gaussian distribution.
Notice that there are two different variances in the system -- $k_n(x_n,x_n)$ and $\sigma^2$. The first variance, $k_n(x_n,x_n)$ is the variance on the estimate induced by the covariance of the input features as defined by a covariance function. $\sigma^2$ is the noise variance of the data as a whole. 
\begin{figure}
\centering
  \epsfig{trim=1.5cm 0 2cm 0,width=0.45\textwidth,file=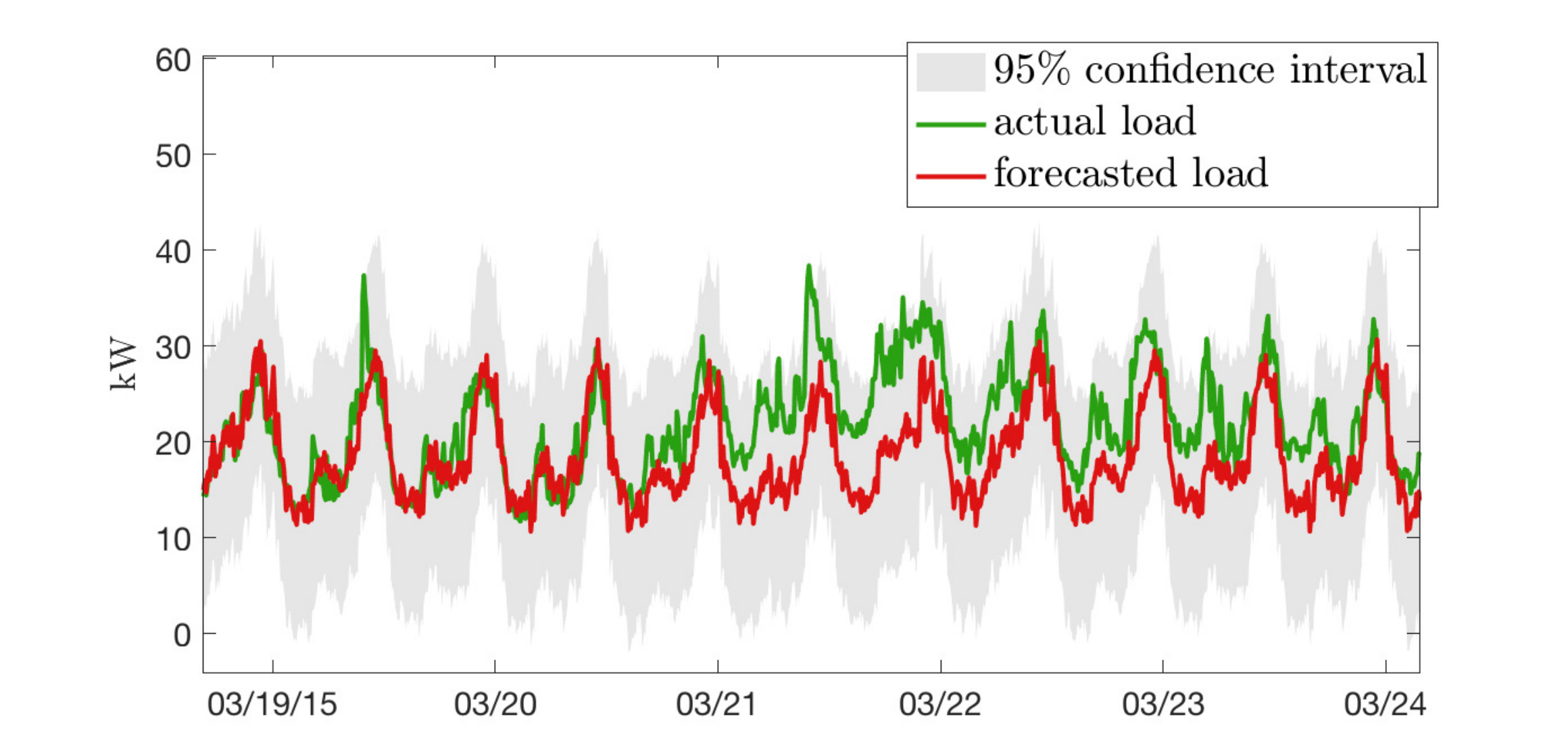}
    \caption{Forecast of an aggregate load using a Gaussian Process model with only discrete-valued time features. Only 10\% of the loads in the aggregate were recorded in historical data. The other 90\% of loads were imputed with the average load profile. Poor forecast performance, such as on March 21st, motivates the use of Bayesian estimation.}
    \label{fig:forecast}
\end{figure}
To challenge the method, in Section~\ref{sec:results}, we consider a GP model that is based on a poor historical data set and no access to real-valued features. 
Fig.~\ref{fig:forecast} exemplifies the resulting forecast accuracy, motivating the use of Bayesian estimation to account for forecast errors such as those experienced on March 21st.



\label{sec:forecasting}

\section{Power Flow Modeling} 
\label{sec:pfm}

Our earlier work~\cite{dobbe_real-time_2016} considered single-phase systems.
Recently, a linear approximations have been proposed for power flow in unbalanced three phase distribution networks~\cite{gan_convex_2014,robbins_optimal_2016,arnold_optimal_2016, sankur_linearized_2016,sankur_optimal_2018}.  This model can be thought of as extensions of the classical \emph{DistFlow} model \cite{baran_optimal_1989} to unbalanced circuits, and was coined the \emph{Dist3Flow} model~\cite{sankur_linearized_2016}.  
In this setting, each node and line segment can have up to three phases, labeled $a$, $b$, and $c$. 
If line $(m,n)$ exists, its phases must be a subset of the phases present at both node $m$ and node $n$.

Let $\mathcal{T} = (\mathcal{N}, \mathcal{E})$ denote a graph representing a radial distribution feeder, where $\mathcal{N}$ is the set of nodes of the feeder and $\mathcal{E}$ is the set of line segments. Nodes are indexed as $n = 0,1,\dots,N$, where node 0 denotes the feeder head (or substation), which we treat as an infinite bus, decoupling interactions in the downstream distribution system from the rest of the grid. 
We also consider a set of nodes equipped with sensors $\mathcal M \subset \mathcal N$.
The current/voltage relationship for a three phase line $(m,n)$ between adjacent nodes $m$ and $n$ is captured by Kirchhoff's Voltage Laws (KVL) in its full \eqref{eq:KVLmn1}, and compact form \eqref{eq:KVLmn2}:
\begin{gather}
	\begin{bmatrix}
		V_{m}^{a} \\ V_{m}^{b} \\ V_{m}^{c}
	\end{bmatrix}
	=
	\begin{bmatrix}
		V_{n}^{a} \\ V_{n}^{b} \\ V_{n}^{c}
	\end{bmatrix}
	+
	\begin{bmatrix}
		Z^{aa}_{mn} & Z^{ab}_{mn} & Z^{ac}_{mn} \\
		Z^{ba}_{mn} & Z^{bb}_{mn} & Z^{bc}_{mn} \\
		Z^{ca}_{mn} & Z^{cb}_{mn} & Z^{cc}_{mn}
	\end{bmatrix}
	\begin{bmatrix}
		I_{mn}^{a} \\ I_{mn}^{b} \\ I_{mn}^{c}
	\end{bmatrix}
    \label{eq:KVLmn1}
	\\
    \mathbf{V}_{m} = \mathbf{V}_{n} + \mathbf{Z}_{mn} \mathbf{I}_{mn}
    \label{eq:KVLmn2}
\end{gather}
Here, $Z^{ab}_{mn} = r^{ab}_{mn} + jx^{ab}_{mn}$ denotes the complex impedance of line $(m,n)$ across phases $a$ and $b$. 

\noindent Next, we define 
the $3 \times 1$ vector of complex power phasors $\mathbf{S}_{mn} = \mathbf{V}_{n} \circ \mathbf{I}_{mn}^{*}$ where $\mathbf{S}_{mn}$ is the power from node $m$ to node $n$ at node $n$.
\begin{equation}
    \sum_{l:(l,m) \in \mathcal{E}} \mathbf{S}_{lm} = \mathbf{s}_{m} + \sum_{n:(m,n) \in \mathcal{E}} \mathbf{S}_{mn} + \mathbf{L}_{mn}
    \label{eq:power03}
\end{equation}
The term $\mathbf{L}_{mn} \in \mathbf{C}^{3 \times 1}$ is a nonlinear and non-convex loss term. 
As in \cite{gan_convex_2014} and \cite{baran_optimal_1989}, we assume that losses are negligible compared to line flows, so that $\left| L_{mn}^{a} \right| \ll \left| S_{mn}^{a} \right| \ \forall (m,n) \in \mathcal{E}$ and all phases. 
Thus, we neglect line losses, linearizing \eqref{eq:power03} into \eqref{eq:power04}.
\begin{equation}
    \sum_{l:(l,m) \in \mathcal{E}} \mathbf{S}_{lm} \approx \mathbf{s}_{m} + \sum_{n:(m,n) \in \mathcal{E}} \mathbf{S}_{mn}
    \label{eq:power04}
\end{equation}
\noindent Now, we define the real scalar $y_{m}^{a} = \left| V_{m}^{a} \right|^{2} = V_{m}^{a} (V_{m}^{a})^{*}$, the $3 \times 1$  real vector $\mathbf{y}_{m} = \left[ y_{m}^{a} , \ y_{m}^{b} , \ y_{m}^{c} \right]^{T} = \mathbf{V}_{m} \circ \mathbf{V}_{m}^{*}$.
With these definitions, \cite{sankur_optimal_2018} derives the following equations that govern the relationship between squared voltage magnitudes and complex power flow across line $(m,n)$:
\begin{equation}
\begin{gathered}
	\mathbf{y}_{m} = \mathbf{y}_{n} + 2 \mathbf{M}_{mn} \mathbf{P}_{mn} - 2 \mathbf{N}_{mn} \mathbf{Q}_{mn} + \mathbf{H}_{mn} \\
    \mathbf{M}_{mn} = \Re \left\{ \Gamma_{n} \circ \mathbf{Z}_{mn}^{*} \right\},
    \mathbf{N}_{mn} = \Im \left\{ \Gamma_{n} \circ \mathbf{Z}_{mn}^{*} \right\} \,,
\end{gathered}
\label{eq:mag09}
\end{equation}
where $\Gamma_{n} = \mathbf{V}_{n} \left( 1 / \mathbf{V}_{n} \right)^{T} \in \mathbb{C}^{3 \times 3}$ represents a matrix with voltage balance ratios across all phases at node $n$. Hence, we have that $\Gamma_{n}\left( a, a \right) = 1$ for the same phase and  $\Gamma_{n}\left( a, b \right) = V_n^{a}/V_n^{b} \triangleq \gamma_{n}^{\phi \psi}$ across phases. Furthermore, $\mathbf{H}_{mn} = \left( \mathbf{Z}_{mn} \mathbf{I}_{mn} \right) \circ \left( \mathbf{Z}_{mn} \mathbf{I}_{mn} \right)^{*} = \left( \mathbf{V}_{m} - \mathbf{V}_{n} \right) \circ \left( \mathbf{V}_{m} - \mathbf{V}_{n} \right)^{*}$ is a $3 \times 1$ real-valued vector representing higher-order terms. 
Notice that we have separated the complex power vector into its active and reactive components, $\mathbf{S}_{mn} = \mathbf{P}_{mn} + j \mathbf{Q}_{mn}$.

This nonlinear and nonconvex system is difficult to incorporate into a state estimation or optimization formulation without the use of convex relaxations. Following the analysis in \cite{gan_convex_2014}, we apply two approximations. The first is that the higher order term $\mathbf{H}_{mn}$, which is the change in voltage associated with losses, is negligible, such that $\mathbf{H}_{mn} \approx \left[ 0, \ 0, \ 0 \right]^{T} \ \forall (m,n) \in \mathcal{E}$. The second assumes that node voltages are ``nearly balanced" (i.e. approximately equal in magnitude and 120$\degree$ apart). This is only applied to $\Gamma_n$ in the RHS of \eqref{eq:mag09}, such that $\gamma_{n}^{ab} = \gamma_{n}^{bc} = \gamma_{n}^{ca} \approx \alpha$, and $\gamma_{n}^{ac} = \gamma_{n}^{ba} = \gamma_{n}^{cb} \approx \alpha^{2}$ for all $n \in \mathcal{N}$.  Under these assumptions, we retrieve
%
\begin{equation}
	\Gamma_{n}
    =
    \begin{bmatrix}
    	1 & \gamma_{n}^{ab} & \gamma_{n}^{ab} \\
        \gamma_{n}^{ba} & 1 & \gamma_{n}^{bc} \\
        \gamma_{n}^{ca} & \gamma_{n}^{cb} & 1
    \end{bmatrix}
    =
    \begin{bmatrix}
    	1 & \alpha & \alpha^{2} \\
        \alpha^{2} & 1 & \alpha \\
        \alpha & \alpha^{2} & 1
    \end{bmatrix}
	\forall n \in \mathcal{N} \,,
    \label{eq:gammaalpha}
\end{equation}
\noindent where $\alpha = 1 \angle 120 \degree = \frac{1}{2} ( -1 + j \sqrt{3} )$ and $\alpha^{-1} = \alpha^{2} = \alpha^{*} = 1 \angle 240 \degree = \frac{1}{2} ( -1 - j \sqrt{3} )$.
Note that we make the ``nearly balanced'' assumption in the process of the formal derivation as in \cite{gan_convex_2014,robbins_optimal_2016}, which does not imply that node voltages need to actually be perfectly balanced for the linearization to be valid. Forthcoming work of the authors shows that for unbalanced networks of medium size, the modeling errors due to linearization and the balanced assumption are below 1\% for power flow scenarios within the power rating of feeders.
Applying the approximations for $\mathbf{H}_{mn}$ and $\Gamma_{n}$ to \eqref{eq:mag09}, we arrive at a linear system of equations:
\begin{equation}
	\mathbf{y}_{m} \approx \mathbf{y}_{n} + 2 \mathbf{M}_{mn} \mathbf{P}_{mn} - 2 \mathbf{N}_{mn} \mathbf{Q}_{mn} \,, \text{ with}
    \label{eq:mag10}
\end{equation}
\begin{equation}
\begin{aligned}
	2 & \mathbf{M}_{mn}
    = \ldots \\
	& \begin{bmatrix}
		2 r_{mn}^{aa} & -r_{mn}^{ab} + \sqrt{3} x_{mn}^{ab} & -r_{mn}^{ac} - \sqrt{3} x_{mn}^{ac} \\
		-r_{mn}^{ba} - \sqrt{3} x_{mn}^{ba} & 2 r_{mn}^{bb} & -r_{mn}^{bc} + \sqrt{3} x_{mn}^{bc} \\
		-r_{mn}^{ca} + \sqrt{3} x_{mn}^{ca} & -r_{mn}^{cb} - \sqrt{3} x_{mn}^{cb} & 2 r_{mn}^{cc}
	\end{bmatrix} \,,
\end{aligned}
\label{eq:Mmn}
\end{equation}
\begin{equation}
\begin{aligned}
	2 & \mathbf{N}_{mn}
    = \ldots \\
	& \begin{bmatrix}
		-2 x_{mn}^{aa} & x_{mn}^{ab} + \sqrt{3} r_{mn}^{ab} & x_{mn}^{ac} - \sqrt{3} r_{mn}^{ac} \\
		x_{mn}^{ba} -\sqrt{3} r_{mn}^{ba} & -2 x_{mn}^{bb} & x_{bc} + \sqrt{3} r_{mn}^{bc}\\
		x_{mn}^{ca} + \sqrt{3} r_{mn}^{ca} & x_{mn}^{cb} -\sqrt{3} r_{mn}^{cb} & -2 x_{mn}^{cc}
	\end{bmatrix} \,.
\end{aligned}
\label{eq:Nmn}
\end{equation}
This linear approximation also enables a linear mapping for voltage angles, similar to~\eqref{eq:mag10}. In the rest of this paper, we will focus on voltage magnitude and leave the extension to voltage angles as an exercise.
In~\cite{sankur_optimal_2018}, the above approximation is assessed for its accuracy. Numerically, for normal operating conditions, the model achieves voltage magnitude errors under 0.5\%, Voltage angle errors under 0.25$^{\circ}$, and line apparent power errors of under 2\%, across the network. It was further observed that these errors increase monotonically with substation power. 


\label{sec:modeling}

\section{Real-time Estimation}

In this Section, we construct the state estimator based on linear least squares estimation. 
This method takes in a prior distribution on measured and unmeasured voltage variables, and updates this in real-time with a limited set of measurements.
To do so, we require the prior statistics of the voltage based on load forecasts (Section~\ref{sec:forecasting}) and power flow modeling (Section~\ref{sec:modeling}). 
We first express measured and unmeasured voltage variables as a linear function of the net load.
We can then construct the necessary matrices to express the voltage forecast as function of load statistics.

\subsection{Voltage as a Function of Net Load}

Consider the vector with all the differences in squared voltage magnitude stacked with the differences in voltage angles over all the branches (i.e. for every set of adjacent nodes) in the network, $\Delta \pmb{y} \in \R^{3 N}$.
%
%
\noindent With~\eqref{eq:mag10}, we can build a model for all the voltage differences over wires throughout the network,
\begin{equation}
\renewcommand{\arraystretch}{1.4}
\Delta \pmb{y}
=
2
\left[
\text{blkdiag}(\mathbf M_{ij}) \ \text{blkdiag}(\mathbf N_{ij})
\right]
\left[
\begin{array}{c}
\text{vec}(\Pb_{ij})  \\
\text{vec}(\Q_{ij})  \\
\end{array}
\right]  = \pmb{Z}_b \pmb{S} \,,
\label{eq:DeltaBranch}
\end{equation}
\noindent where $\Scap \in \R^{6N}$ is the vector with real and reactive branch flows stacked vertically, and $\pmb{Z}_b \in \R^{3N \times 6N}$ is a horizontal stack of two block diagonal matrices with the corresponding 3-by-3 matrices from respectively~\eqref{eq:Mmn} and~\eqref{eq:Nmn}.
With~\eqref{eq:power04}, we can express the branch flows $\Scap$ in terms of the nodal net loads, which yields $\Scap = \mathcal P_b \pmb{s}$,
%
%
with $\s \in \R^{6N}$ a vector with the nodal net loads, real and reactive power $p_n,q_n\,, \, n \in \N$ stacked vertically, and $\mathcal P_b \in \R^{6N \times 6N}$ a binary matrix in which a row represents a branch with 1s selecting the nodes downstream of the branch. 
We have now expressed the differences in voltage magnitude over all $N$ lines in terms of the nodal load vector,
\begin{equation}
\Delta \pmb{y}  = \pmb{Z}_b \mathcal P_b \s \triangleq \pmb{Z}_n \s \,,
\label{eq:loadeqns}
\end{equation}
where $\pmb{Z}_n \triangleq \pmb{Z}_b \mathcal P_b \in \R^{3N \times 6N}$.

\subsubsection{Measured quantities}
In our actual setting, we do not directly measure voltage differences over all individual wires. Instead, we place the sensors over a distance spanning multiple branches and buses.
The voltage difference over the path can be rewritten as the sum of the individual differences of the branches lying on the path, 
\begin{equation}
\Delta \pmb{y}_{m} \triangleq
\left[
\begin{array}{c}
\y_{m_2} - \y_{m_1} \\
\vdots \\
\y_{m_{M}} - \y_{m_{M-1}} \\
\end{array}
\right] \in \R^{3(M-1)} \,,
\end{equation}
with $m_1, \hdots, m_M \in \mathcal M$. We can now formulate the equations, by adding up the differences of all individual lines in between the sensors, by formulating a permutation matrix such that $\Delta \pmb{y}_{m} = \mathcal P_{m} \Delta \pmb{y}$, and hence
\begin{equation}
\Delta \pmb{y}_{m} = \mathcal P_{m} \pmb{Z}_n \s = \pmb{Z}_{m} \s \,,
\label{eq:measeqns}
\end{equation}
where $\pmb{Z}_{m} \triangleq \mathcal P_{m} \pmb{Z}_n = \mathcal P_{m} \pmb{Z}_b \mathcal P_b \in \R^{3(M-1) \times 6N}$. 
This gives us an expression for the measured quantities as a function of the nodal load vector.

\subsubsection{Non-measured quantities - Voltage Estimation}

We are interested to estimate voltage magnitude and angle at all the $N-M$ buses in the network that are not equipped with a sensor. We aim to do this given a measurement of the voltage phasor at a limited number of $M$ buses in the network, and forecast statistics on the nodal load vector $\s$.
We consider the differences in voltage between a location we want to estimate and a nearby sensor location. These differences are collected in a vector $\Delta \pmb{Y}_e$ to be estimated as a function of the load vector $\s$, similar to the construction of the measurement equation:
\begin{equation}
\Delta \pmb{y}_e = \pmb{Z}_e \s \in \R^{3(N + 1 - M)} \,,
\end{equation}
where $\pmb{Z}_e \triangleq \mathcal P_{m} \pmb{Z}_n = \mathcal P_{e} \pmb{Z}_b \mathcal P_b \in \R^{3(N + 1 - M) \times 6N}$ is constructed in the same way as $\pmb{Z}_m$ in \eqref{eq:measeqns}. In order to retrieve an estimate of the absolute voltage value, we can simply take the nearest sensor reading and add/subtract the estimated difference between the location and that sensor location.

\subsection{Voltage Forecast Statistics}

We now have that our measurements are voltage phasor differences, i.e. $z = \Delta \pmb{y}_m$ and the estimation quantities are other voltage phasor difference, i.e. $x = \Delta \pmb{y}_e$. Given the linear relationships with the load vector $\s$, we can now derive the statistics on $z$. The mean of $z$ is
\begin{equation}
\mu_z (t) = E(\Delta \pmb{y}_{m}) = \pmb{Z}_{m} \mu_s(t) \,.
\end{equation}
Similarly, we have that $\mu_x(t) =  E(\Delta \pmb{y}_e) =  Z_e \mu_s(t)$. This quantity is already useful as a forecast in operation (available before any real-time measurements are gathered).
The covariance of $z$ is 
\begin{equation}
\Sigma_z (t) = E((z - \mu_z)(z - \mu_z)^{\top}) = \pmb{Z}_{m} \Sigma_s(t) \pmb{Z}_{m}^{\top} \,. 
\end{equation}
Similarly, we have that the cross-covariance of $x$ and $z$ is $\Sigma_{x,z} (t) = \pmb{Z}_e \Sigma_s(t) \pmb{Z}_{m}^{\top}$.
This yields all the statistics we need to construct the distribution grid state estimator. 

\subsection{Constructing the State Estimator}
We can now analytically derive the LLSE of $\pmb{y}_e$ given measurements $\pmb{y}_m$, as a specific form of~\eqref{eq:llse} presented in Section~\ref{subsec:mmse}. 
For our voltage estimation setting this yields
\begin{equation}
\begin{array}{rl}
L[\Delta \pmb{y}_e | \Delta \pmb{y}_m ] &= E(\Delta \pmb{y}_e) + \hdots \\ 
&\Sigma_{\Delta \pmb{y}_e,\Delta \pmb{y}_m} \Sigma^{-1}_{\Delta \pmb{y}_m} (\Delta \pmb{y}_m - E(\Delta \pmb{y}_m )) \,, \\
&=  \pmb{Z}_{e} \mu_s + \hdots \\
& \pmb{Z}_{e} \Sigma_s \pmb{Z}_{m}^{\top} \left(  \pmb{Z}_{m} \Sigma_s \pmb{Z}_{m}^{\top} \right)^{-1} (\Delta \pmb{y}_m - \pmb{Z}_{m} \mu_s) \,,
\end{array}
\label{eq:llse_v}
\end{equation}
where we dropped the time index for brevity.
With $L[\Delta \pmb{y}_e | \Delta \pmb{y}_m ]$, the voltage estimates can be retrieved as
\begin{equation}
    \hat{\pmb{V}}_e = \sqrt{\pmb{y}_{\text{near}} + L[\Delta \pmb{y}_e | \Delta \pmb{y}_m ]} \,,
\label{eq:llse_voltage}
\end{equation}
where $\pmb{V}_e$ denotes a stacked vector with voltages for all buses without measurement, and $\pmb{y}_{\text{near}}$ are the squared voltages at the nearest measured bus for each estimated bus.
\eqref{eq:llse_v} is written in the form $\Delta \hat{\pmb{y}}_e = f(\Delta \pmb{y}_m)$; apart from the measurements, other information needed to evaluate the estimator are forecast statistics, available \emph{a priori}. As such, \eqref{eq:llse_v} uses the statistical information of the net loads $\s$, in combination with the topology and impedance information of the network, to present a closed-form analytical estimator of voltage differences throughout the network, which is \emph{linear} in the measurements and computed efficiently in real-time.

Note that in order for the inversion in \eqref{eq:llse_v} to be feasible we need $\pmb{Z}_{m}$ to be full row rank, which means $\mathcal P_{m}$ needs to be full row rank. In practice, this condition holds if the sensors are placed in such a way that the paths between sensors and their closest upstream sensors are not fully overlapping. 

Lastly, the method has the potential to detect and account for unregular conditions, such as bad data or faults. The estimator compares forecasted voltage differences between two sensors to the actual voltage differences, leading to an ‘innovation’ that triggers an update of the unmeasured voltages. If this comparison indicates that the measured voltage difference between two sensors is unusually large, for instance too large to represent a realistic power flow scenario, this may indicate bad data or a fault scenario. One may either prevent such data from entering the estimation, or use this feature of the estimation step towards fault detection. A further analysis of bad data and fault detection estimation is left as future work.

\label{sec:estimation}

\section{Results}
\label{sec:results}
We test the method on three systems; a proof-of-concept on the simplified 37 node test feeder, a scaling assessment on the 8500 node test feeder, and a validation on a real feeder in operation with 142 nodes.

\subsection{Comparison to conventional WLS method}
\begin{figure*}
\centering
  \includegraphics[trim=5cm 0 5cm 0,width=\textwidth]{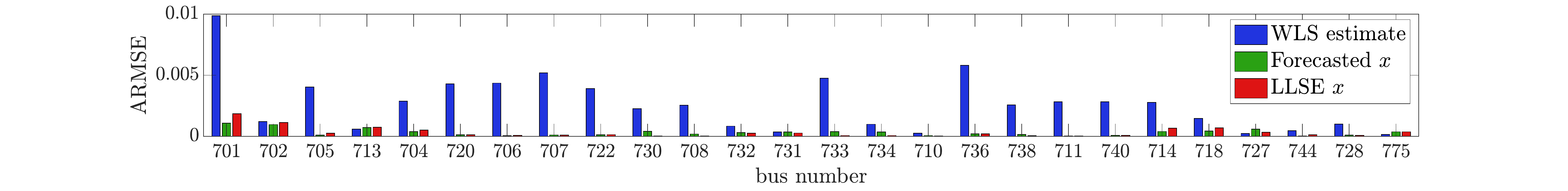}
    \caption{ARMSE in p.u. for each non-measured bus, comparing conventional WLS method, forecasting method and estimation method.}
    \label{fig:IEEE37_ARMSEwithWLS}
\end{figure*}

\begin{figure*}
\centering
  \includegraphics[trim=5cm 0cm 1cm 0cm,width=\textwidth]{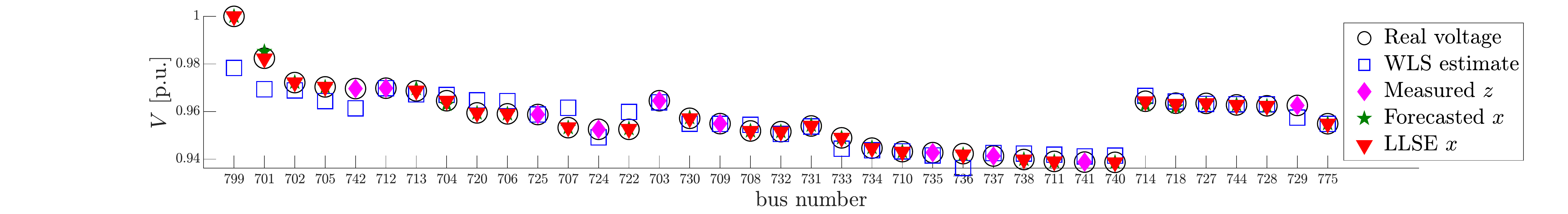}
    \caption{Example voltage profile with forecast and estimation update at all the buses, with comparison to conventional WLS method.}
    \label{fig:IEEE37_VoltagewithWLS}
\end{figure*}

\begin{figure}
\centering
  \includegraphics[trim=2cm 0cm 2cm 0cm,width=0.4\textwidth,]{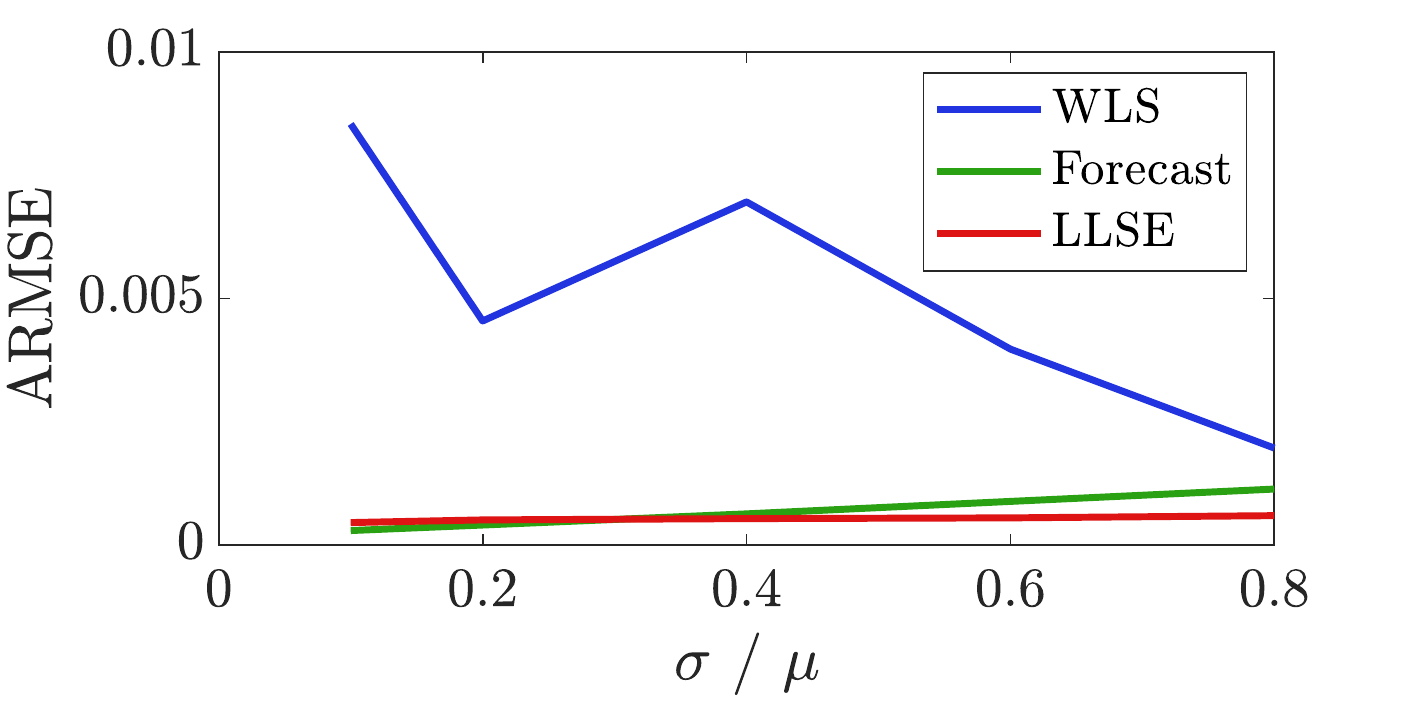}
    \caption{Comparing methods for increasing uncertainty in the forecast statistics. ARMSE is computed across all estimated buses, across phases and simulations, for the sensor placement indicated by the magenta diamonds in Figure~\ref{fig:IEEE37_VoltagewithWLS}.}
    \label{fig:comparisonoversigma}
\end{figure}

Earlier work implemented the distribution grid state estimator on a single-phase radial network~\cite{dobbe_real-time_2016}. In Figures~\ref{fig:IEEE37_ARMSEwithWLS}, we compare these results to a traditional WLS implementation with a flat start (plus some Gaussian noise with $\sigma = 0.05$ p.u.)~\cite{abur_power_2004}. 
To make a fair comparison, we allow the WLS to have the same 10 PMU sensors, complemented with pseudomeasurements comprising forecasts of bus injections or branch flows.
To achieve observability, we carefully construct the measurements using the orthogonal transformation as proposed in~\cite{monticelli_observability_1986}.
We see that across all buses, our forecast (constructed with $E(\Delta \pmb{y}_e) = \pmb{Z}_{e} \mu_s $ as in~\eqref{eq:llse_v}) and the LLSE results based on \eqref{eq:llse_voltage} either roughly match or outperform the conventional WLS.
Figure~\ref{fig:IEEE37_VoltagewithWLS} shows an instance of a voltage profile with all compared methods.
Figure~\ref{fig:comparisonoversigma} compares the methods for increasing levels of variance in the forecast statistics, assigning the variance as a fixed ratio of each load forecast mean. As expected due to their linearity, the forecast and LLSE method both show linearly increasing ARMSE errors for increasing variance, with the LLSE method performing roughly 50\% better than the forecast. WLS performs worse overall; the intrinsic error of the optimization overshadow the impact of forecast uncertainty. 

\subsection{Three-phase experiments on 37 and 8500 Node Test Feeders}

\begin{figure*}
\centering
  \includegraphics[trim=3cm 0 3cm 0,width=\textwidth]{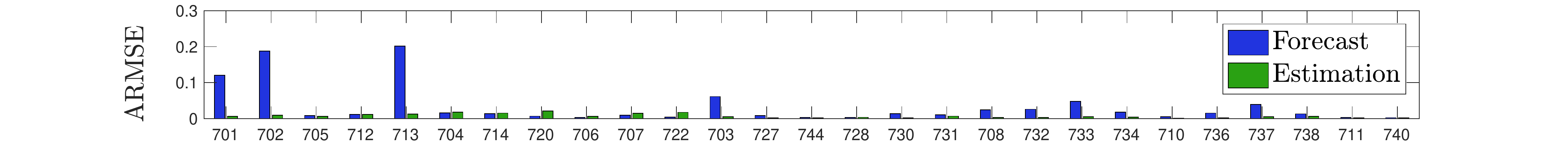}
    \caption{ARMSE in p.u. for each non-measured bus across all phases. Buses with higher forecast errors benefit significantly from the estimation procedure. Buses that already have good accuracy on forecasted values of the order $< 0.01$ p.u. do not necessarily gain much from estimation.}
    \label{fig:IEEE37_ARMSE}
\end{figure*}
\begin{figure*}
\centering
  \includegraphics[trim=0cm 0cm 0cm 0cm,width=\textwidth]{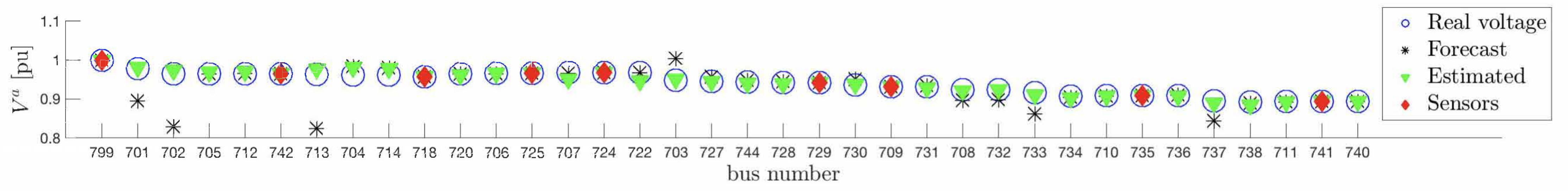}
    \caption{Example voltage profile with forecast and estimation update at all the buses for phase $a$.}
    \label{fig:IEEE37_Voltage}
\end{figure*}

We use the linear model presented in Section~\ref{sec:pfm} to extend our results to three-phase systems. We first implement the method on the 37 Node Test Feeder. 
We include all capacitor banks and voltage regulators and assume all loads are constant power. 
As load data, we use data sets provided by Pecan Street for educational use~\cite{noauthor_dataport_2017}. 
The raw data contains 15-minute-interval data sampled from July 1, 2013 to September 26, 2016. 
In this paper, a time $t$ represents the number of 15 minute intervals starting from July 1, 2013, or $t_0$, where $t_0 = 0$. 
We aggregate different household time series from the Pecan Street data set such that the aggregated time series data had a maximum equivalent to the spot loads defined by the IEEE feeder model~\cite{noauthor_ieee_2017}. 
The aggregated time series are then used to build a Gaussian Process forecast model for real and reactive power at each bus, as outlined in Section \ref{sec:forecasting}.
Voltage sensors are placed at 9 different buses, indicated by red diamonds in Fig.~\ref{fig:IEEE37_Voltage}. 
To assess performance, we compute the Average Root Mean Square Error (ARMSE) on the voltages $\pmb{V}_e$ that are not measured by PMUs and thus estimated,
\begin{equation}
    \text{ARMSE}(\{\hat{\pmb{V}}_e[t]\}_{t=1}^T) =  \sqrt{\frac{1}{T} \displaystyle \sum_{t=1}^T \| \hat{\pmb{V}}_e[t] - \pmb{V}_e[t]  \|^2} \,.
\end{equation}
Fig.~\ref{fig:IEEE37_ARMSE} shows the ARMSE metric for all buses. It is bounded by 0.2 p.u. for the forecasted values and 0.02 p.u. for the estimated values. Notice that buses with higher forecast errors benefit significantly from the estimation procedure. Buses that have higher forecast accuracy of the order $< 0.01$ p.u. do not gain much from estimation. This can be attributed to the errors being of the same order as modeling errors due to linear approximation, which are carefully studied in a separate forthcoming paper~\cite{sankur_optimal_2018}.
Fig.~\ref{fig:IEEE37_Voltage} shows a representative result for one power flow instance, with the true voltage indicated by blue circles. Although forecasted voltage values (black stars) are close for most buses, there are some buses where large errors arise. These are effectively addressed by the estimation scheme (green triangles).

The average computation time for this feeder with 9 sensors is less than 0.015 $\mu$s for an implementation on Matlab 2017b on a Macbook Pro with 2.8 GHz processor and 16GB 2133Mhz memory.
%
To understand the performance of the method for large scale networks, we apply the method to the 8500 Node Test Feeder, which resembles the large scale and complexity of modern distribution utility planning model for a circuit~\cite{schneider_analytic_2018}. The feeder voltage and power ratings were left unchanged (7.2 kV and 2.5 MVA), as were line segment configuration assignments. We include all capacitor banks and voltage regulators, and focus our simulations to the primary side of the network, comprising 2518 nodes.
This network has 68 times as many nodes as the 37 node test feeder, and to make a fair scaling comparison in terms of computational time, we also install 68 times as many sensors, equalling 612. The average computation time grows linearly to 1 $\mu$s. 
In terms of accuracy, we retrieve satisfactory results for the forecasted voltage values, validating the use of the linear approximation on larger networks. However, the accuracy of the estimated values has so far not been consistent. Hence, the application of the LLSE method to larger networks requires further analysis. Next, we consider a network of moderate size with 142 nodes. 
\label{sec:simulations}

\subsection{Validation On A Utility Testbed}


\begin{figure}
\centering
  \epsfig{width=0.48\textwidth,file=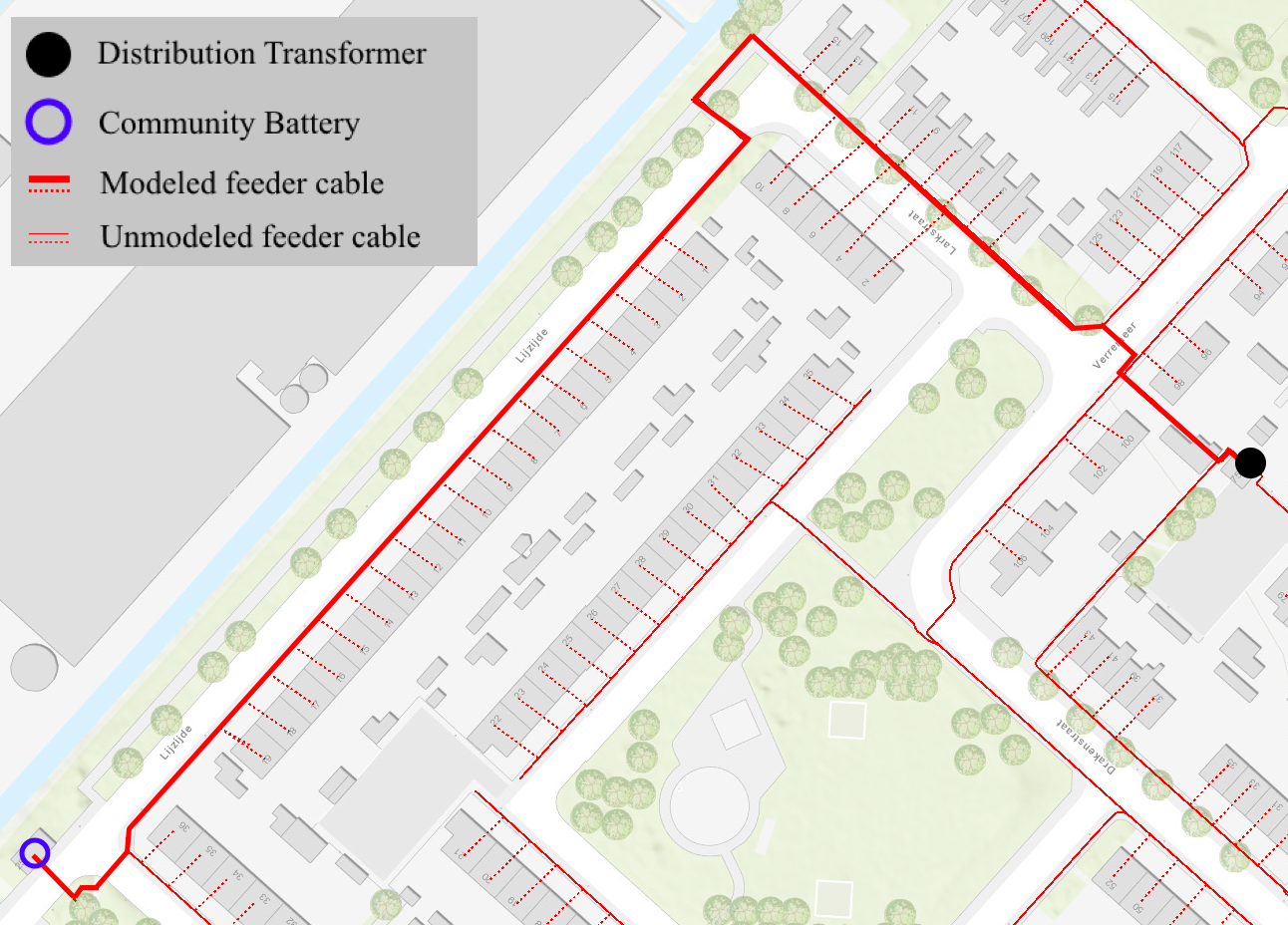}
    \caption{Alliander's low voltage network of Rijsenhout.}
    \label{fig:Rijsenhout2}
\end{figure}

We validate the method on a network in the territory of \emph{Alliander}, the largest Distribution Network Operator (DNO) of the Netherlands serving over three million customers.
Alliander is experimenting with community electricity storage in Rijsenhout, a suburban village close to Amsterdam. The project is titled ``BuurtBatterij'' which translates to ``Community Battery''. Fig.~\ref{fig:Rijsenhout2} depicts the Rijsenhout feeder that houses the battery project. 
As a part of the community battery experiments, the local low voltage power grid is modeled and load and voltage data are gathered. 


The feeder contains 142 buses, of which 34 are regular household customers, one is the distribution transformer and one is the community battery. The other buses are network cable joints.
The source of the network data is the Alliander GIS database, which contains the exact location and properties of the electricity cables. However, the GIS database  does not contain on which phase each customer is connected, therefore the estimator is constructed using a balanced single phase model, using the formulation in~\cite{dobbe_real-time_2016}. 
The distribution transformer is located at the top of the feeder, and the Neighborhood Battery is installed at the end of the feeder. 
Both the transformer and battery contain SCADA equipment for measuring power and voltage at a 1-second rate. 
Of the 34 households connected to this feeder, 12 customers share their power consumption data with Alliander as part of the community battery project. 
All data for building forecasts have been collected at a 1-minute resolution.
Customers with no direct measurement were assigned the residual power load, which was defined as the total transformer load minus the sum of all measured loads. Each unmeasured customer was assigned an equal proportional share of the residual load. Note that this introduces bias in the voltage forecasts.

\begin{figure*}
    \centering
    \epsfig{trim=4cm 0 4cm 0,width=\textwidth,file=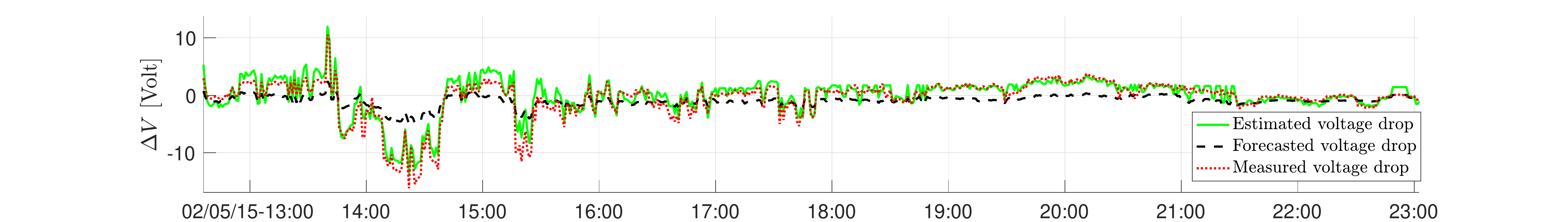}
    \caption{Comparison of forecasted and estimated voltage with real voltage measurement at the Neighborhood Battery bus.}
    \label{fig:BBvalidation}
\end{figure*}

Fig.~\ref{fig:BBvalidation} compares the forecasted and estimated voltage drop at a particular bus with real voltage measurements. 
The estimated values provide a significant improvement over the forecasted values, showing agreement with the actual values.
The improvements are stronger for larger voltage deviations, providing critical information for safety procedures.
At certain times the estimation does not improve accuracy, which has two explanations. 
Firstly, for smaller voltage deviations, modeling errors due to linearization of power flow are more dominant, as mentioned above.
Secondly, the effect of limited real-time voltage sensors (in this case only 2 out of 140 buses) provides significant but limited improvement due to limited observability of all load flow scenarios in the network. This challenge requires revisiting the notion of network observability, which is a topic of future research.
The estimation significantly reduces the ARMSE across all buses in the network, on average by 60\%.
Given the difficulty of predicting the power consumption of individual househoulds due to their variability, this result is useful for DSOs in improving the fidelity of their forecasting data with limited sensing capabilities, which is a likely context in most networks for the foreseeable future.
Given these results, Alliander plans to use the method for optimal sensor placement, cable health monitoring, real time overload predictions, and control of voltage and power flow.
\label{sec:validation}



\section{Conclusions}

This paper addressed the challenge of formulating an estimator for voltages, for scenarios where
fully observed sensor arrangements are not yet feasible, and load forecasts may be subject to large uncertainties.
Building on preliminary work for single-phase networks, we developed a method for three-phase networks that exploits new linear approximations to construct statistics of relevant voltage variables based on load statistics. 
We then used a Bayesian approach, in the form of the linear least squares estimator, to update prior voltage statistics in real-time based on measured deviations at a limited set of voltage sensors, at minimal computational cost.
We applied the method to test feeders and validated it on a real testbed, showing its ability to provide useful voltages estimates using limited historical data and real-time sensors.
As such, the method is highly applicable in the typical distribution network setting in which ubiquitous sensing will remain limited for the foreseeable future.
\label{sec:conclusions}





\section*{Acknowledgments}
We thank Michael Sankur for contributions to the code base for this work, Sascha von Meier for helpful suggestions in the earlier stage of this research, and colleagues at Alliander for enabling the experiments for real world integration. This work is supported by two US National Science Foundation Grants: CPS FORCES (award number CNS-1239166) and CyberSEES (award number 1539585), and by the UC-Philippine-California Advanced Research Institute (award number IIID-2015-10).


\ifCLASSOPTIONcaptionsoff
  \newpage
\fi



%
 \bibliographystyle{IEEEtran}
 \bibliography{references.bib} 

\begin{thebibliography}{10}
\providecommand{\url}[1]{#1}
\csname url@samestyle\endcsname
\providecommand{\newblock}{\relax}
\providecommand{\bibinfo}[2]{#2}
\providecommand{\BIBentrySTDinterwordspacing}{\spaceskip=0pt\relax}
\providecommand{\BIBentryALTinterwordstretchfactor}{4}
\providecommand{\BIBentryALTinterwordspacing}{\spaceskip=\fontdimen2\font plus
\BIBentryALTinterwordstretchfactor\fontdimen3\font minus
  \fontdimen4\font\relax}
\providecommand{\BIBforeignlanguage}[2]{{%
\expandafter\ifx\csname l@#1\endcsname\relax
\typeout{** WARNING: IEEEtran.bst: No hyphenation pattern has been}%
\typeout{** loaded for the language `#1'. Using the pattern for}%
\typeout{** the default language instead.}%
\else
\language=\csname l@#1\endcsname
\fi
#2}}
\providecommand{\BIBdecl}{\relax}
\BIBdecl

\bibitem{kaur_effects_2006}
G.~Kaur and M.~Vaziri, ``Effects of distributed generation ({DG})
  interconnections on protection of distribution feeders,'' in \emph{{IEEE}
  {Power} {Engineering} {Society} {General} {Meeting}, 2006}, 2006.

\bibitem{seltenrich_new_2013}
\BIBentryALTinterwordspacing
N.~Seltenrich, ``The {New} {Grid} - {Plugging} into {California}'s clean-energy
  future,'' 2013. [Online]. Available:
  \url{https://nature.berkeley.edu/breakthroughs/fa12/the_new_grid}
\BIBentrySTDinterwordspacing

\bibitem{zhao_short-term_2017}
J.~Zhao, G.~Zhang, M.~La~Scala, Z.~Y. Dong, C.~Chen, and J.~Wang, ``Short-term
  state forecasting-aided method for detection of smart grid general false data
  injection attacks,'' \emph{IEEE Transactions on Smart Grid}, vol.~8, no.~4,
  pp. 1580--1590, 2017.

\bibitem{abur_power_2004}
\BIBentryALTinterwordspacing
A.~Abur and A.~G{\textbackslash}'omez~Exp\{{\textbackslash}'o\}sito,
  \emph{\BIBforeignlanguage{en}{Power {System} {State} {Estimation}: {Theory}
  and {Implementation}}}, ser. Power {Engineering} ({Willis}).\hskip 1em plus
  0.5em minus 0.4em\relax CRC Press, Mar. 2004, vol.~24. [Online]. Available:
  \url{http://www.crcnetbase.com/doi/book/10.1201/9780203913673}
\BIBentrySTDinterwordspacing

\bibitem{giannakis_monitoring_2013}
G.~Giannakis, V.~Kekatos, N.~Gatsis, S.-J. Kim, H.~Zhu, and B.~Wollenberg,
  ``Monitoring and {Optimization} for {Power} {Grids}: {A} {Signal}
  {Processing} {Perspective},'' \emph{IEEE Signal Processing Magazine},
  vol.~30, no.~5, pp. 107--128, Sep. 2013.

\bibitem{monticelli_network_1985-1}
\BIBentryALTinterwordspacing
A.~Monticelli and F.~F. Wu, ``Network {Observability}: {Theory},'' \emph{IEEE
  Transactions on Power Apparatus and Systems}, vol. PAS-104, no.~5, pp.
  1042--1048, May 1985. [Online]. Available:
  \url{http://ieeexplore.ieee.org/document/4118842/}
\BIBentrySTDinterwordspacing

\bibitem{korres_optimal_2015}
G.~N. Korres, N.~M. Manousakis, T.~C. Xygkis, and J.~Lofberg, ``Optimal phasor
  measurement unit placement for numerical observability in the presence of
  conventional measurements using semi-definite programming,''
  \emph{Transmission Distribution IET Generation}, vol.~9, no.~15, pp.
  2427--2436, 2015.

\bibitem{von_meier_micro-synchrophasors_2014}
\BIBentryALTinterwordspacing
A.~von Meier, D.~Culler, A.~McEachern, and R.~Arghandeh, ``Micro-synchrophasors
  for distribution systems,'' in \emph{{IEEE} 5th {Innovative} {Smart} {Grid}
  {Technologies} {Conference}, {Washington}, {DC}}, 2014. [Online]. Available:
  \url{http://pqubepmu.com/ARPA-E_uPMU_Project_Overview_Apr_19_2013.pdf}
\BIBentrySTDinterwordspacing

\bibitem{dobbe_real-time_2016}
\BIBentryALTinterwordspacing
R.~Dobbe, D.~Arnold, S.~Liu, D.~Callaway, and C.~Tomlin, ``Real-time
  distribution grid state estimation with limited sensors and load
  forecasting,'' in \emph{7th {International} {Conference} on
  {Cyber}-{Physical} {Systems} ({ICCPS})}.\hskip 1em plus 0.5em minus
  0.4em\relax ACM/IEEE, May 2016. [Online]. Available:
  \url{http://ieeexplore.ieee.org/abstract/document/7479117/}
\BIBentrySTDinterwordspacing

\bibitem{schenato_bayesian_2014}
L.~Schenato, G.~Barchi, D.~Macii, R.~Arghandeh, K.~Poolla, and A.~Von~Meier,
  ``Bayesian linear state estimation using smart meters and {PMUs} measurements
  in distribution grids,'' in \emph{2014 {IEEE} {International} {Conference} on
  {Smart} {Grid} {Communications} ({SmartGridComm})}, Nov. 2014, pp. 572--577.

\bibitem{weng_probabilistic_2017}
Y.~Weng, R.~Negi, and M.~D. Ilic, ``Probabilistic {Joint} {State} {Estimation}
  for {Operational} {Planning},'' \emph{IEEE Transactions on Smart Grid},
  vol.~PP, no.~99, 2017.

\bibitem{rousseaux_whither_1990}
\BIBentryALTinterwordspacing
P.~Rousseaux, T.~Van~Cutsem, and T.~E. Dy~Liacco, ``Whither dynamic state
  estimation?'' \emph{International Journal of Electrical Power \& Energy
  Systems}, vol.~12, no.~2, pp. 104--116, Apr. 1990. [Online]. Available:
  \url{http://www.sciencedirect.com/science/article/pii/014206159090006W}
\BIBentrySTDinterwordspacing

\bibitem{blood_electric_2008}
E.~Blood, B.~Krogh, and M.~Ilic, ``Electric power system static state
  estimation through {Kalman} filtering and load forecasting.''\hskip 1em plus
  0.5em minus 0.4em\relax IEEE, Jul. 2008.

\bibitem{singh_distribution_2010}
\BIBentryALTinterwordspacing
R.~Singh, B.~C. Pal, and R.~A. Jabr, ``Distribution system state estimation
  through {Gaussian} mixture model of the load as pseudo-measurement,''
  \emph{IET Generation, Transmission \&amp; Distribution}, vol.~4, no.~1, pp.
  50--59, Jan. 2010. [Online]. Available:
  \url{http://digital-library.theiet.org/content/journals/10.1049/iet-gtd.2009.0167}
\BIBentrySTDinterwordspacing

\bibitem{manitsas_distribution_2012}
E.~Manitsas, R.~Singh, B.~C. Pal, and G.~Strbac, ``Distribution {System}
  {State} {Estimation} {Using} an {Artificial} {Neural} {Network} {Approach}
  for {Pseudo} {Measurement} {Modeling},'' \emph{IEEE Transactions on Power
  Systems}, vol.~27, no.~4, pp. 1888--1896, Nov. 2012.

\bibitem{guo_multi-time_2015}
Y.~Guo, B.~Zhang, W.~Wu, and H.~Sun, ``Multi-time interval power system state
  estimation incorporating phasor measurements,'' in \emph{Power \& {Energy}
  {Society} {General} {Meeting}, 2015 {IEEE}}.\hskip 1em plus 0.5em minus
  0.4em\relax IEEE, 2015, pp. 1--5.

\bibitem{zhao_power_2016}
J.~Zhao, G.~Zhang, K.~Das, G.~N. Korres, N.~M. Manousakis, A.~K. Sinha, and
  Z.~He, ``Power system real-time monitoring by using {PMU}-based robust state
  estimation method,'' \emph{IEEE Transactions on Smart Grid}, vol.~7, no.~1,
  pp. 300--309, 2016.

\bibitem{asprou_two-stage_2017}
M.~Asprou, S.~Chakrabarti, and E.~Kyriakides, ``A two-stage state estimator for
  dynamic monitoring of power systems,'' \emph{IEEE Systems Journal}, vol.~11,
  no.~3, pp. 1767--1776, 2017.

\bibitem{manousakis_hybrid_2018}
N.~M. Manousakis and G.~N. Korres, ``A hybrid power system state estimator
  using synchronized and unsynchronized sensors,'' \emph{International
  Transactions on Electrical Energy Systems}, p. e2580, 2018.

\bibitem{da_silva_state_1983}
A.~L. Da~Silva, M.~B. Do~Coutto~Filho, and J.~F. De~Queiroz, ``State
  forecasting in electric power systems,'' in \emph{{IEE} {Proceedings} {C}
  ({Generation}, {Transmission} and {Distribution})}, vol. 130.\hskip 1em plus
  0.5em minus 0.4em\relax IET, 1983, pp. 237--244.

\bibitem{do_coutto_filho_forecasting-aided_2009}
M.~B. Do~Coutto~Filho, J.~C.~S. de~Souza, and R.~S. Freund, ``Forecasting-aided
  state estimation - {Part} {II}: {Implementation},'' \emph{IEEE Transactions
  on Power Systems}, vol.~24, no.~4, pp. 1678--1685, 2009.

\bibitem{primadianto_review_2017}
A.~Primadianto and C.~N. Lu, ``A {Review} on {Distribution} {System} {State}
  {Estimation},'' \emph{IEEE Transactions on Power Systems}, vol.~32, no.~5,
  pp. 3875--3883, Sep. 2017.

\bibitem{gol_hybrid_2015}
M.~G\{{\textbackslash}"\{o\}\}l and A.~Abur, ``A {Hybrid} {State} {Estimator}
  {For} {Systems} {With} {Limited} {Number} of {PMUs},'' \emph{IEEE
  Transactions on Power Systems}, vol.~30, no.~3, pp. 1511--1517, May 2015.

\bibitem{sankur_linearized_2016}
\BIBentryALTinterwordspacing
M.~D. Sankur, R.~Dobbe, E.~Stewart, D.~S. Callaway, and D.~B. Arnold, ``A
  {Linearized} {Power} {Flow} {Model} for {Optimization} in {Unbalanced}
  {Distribution} {Systems},'' \emph{arXiv preprint arXiv:1606.04492}, 2016.
  [Online]. Available:
  \url{https://www.researchgate.net/profile/Emma_Stewart2/publication/303970012_A_Linearized_Power_Flow_Model_for_Optimization_in_Unbalanced_Distribution_Systems/links/578903a208ae7a588ee85778.pdf}
\BIBentrySTDinterwordspacing

\bibitem{sankur_optimal_2018}
\BIBentryALTinterwordspacing
M.~D. Sankur, R.~Dobbe, A.~von Meier, E.~M. Stewart, and D.~B. Arnold,
  ``Optimal {Voltage} {Phasor} {Regulation} for {Switching} {Actions} in
  {Unbalanced} {Distribution} {Systems},'' \emph{arXiv:1804.02080 [math]}, Apr.
  2018, arXiv: 1804.02080. [Online]. Available:
  \url{http://arxiv.org/abs/1804.02080}
\BIBentrySTDinterwordspacing

\bibitem{walrand_probability_2014}
J.~Walrand, \emph{Probability in {Electrical} {Engineering} and {Computer}
  {Science}: {An} {Application}-{Driven} {Course}}, 1st~ed.\hskip 1em plus
  0.5em minus 0.4em\relax Quoi?, Mar. 2014, vol.~1.

\bibitem{kalman_new_1960}
R.~E. Kalman, ``A new approach to linear filtering and prediction problems,''
  \emph{Journal of basic Engineering}, vol.~82, no.~1, pp. 35--45, 1960.

\bibitem{mirowski_demand_2014}
\BIBentryALTinterwordspacing
P.~Mirowski, S.~Chen, T.~Kam~Ho, and C.-N. Yu, ``Demand forecasting in smart
  grids,'' \emph{Bell Labs technical journal}, vol.~18, no.~4, pp. 135--158,
  2014. [Online]. Available:
  \url{http://onlinelibrary.wiley.com/doi/10.1002/bltj.21650/full}
\BIBentrySTDinterwordspacing

\bibitem{huang_short-term_2003}
S.-J. Huang and K.-R. Shih, ``Short-term load forecasting via {ARMA} model
  identification including non-{Gaussian} process considerations,'' \emph{IEEE
  Transactions on Power Systems}, vol.~18, no.~2, pp. 673--679, May 2003.

\bibitem{behl_data-driven_2016}
M.~Behl, A.~Jain, and R.~Mangharam, ``Data-{Driven} {Modeling}, {Control} and
  {Tools} for {Cyber}-{Physical} {Energy} {Systems}.''\hskip 1em plus 0.5em
  minus 0.4em\relax Vienna, Austria: ACM/IEEE, Apr. 2016.

\bibitem{rasmussen_gaussian_2004}
C.~E. Rasmussen, ``Gaussian processes in machine learning,'' in \emph{Advanced
  lectures on machine learning}.\hskip 1em plus 0.5em minus 0.4em\relax
  Springer, 2004, pp. 63--71.

\bibitem{mori_probabilistic_2005}
H.~Mori and M.~Ohmi, ``Probabilistic short-term load forecasting with
  {Gaussian} processes,'' Nov. 2005.

\bibitem{gan_convex_2014}
L.~Gan and S.~H. Low, ``Convex relaxations and linear approximation for optimal
  power flow in multiphase radial networks,'' Aug. 2014.

\bibitem{robbins_optimal_2016}
B.~A. Robbins and A.~D. Dominguez-Garcia, ``Optimal reactive power dispatch for
  voltage regulation in unbalanced distribution systems,'' \emph{IEEE
  Transactions on Power Systems}, vol.~31, no.~4, pp. 2903--2913, 2016.

\bibitem{arnold_optimal_2016}
D.~B. Arnold, M.~Sankur, R.~Dobbe, K.~Brady, D.~S. Callaway, and A.~V. Meier,
  ``Optimal dispatch of reactive power for voltage regulation and balancing in
  unbalanced distribution systems,'' in \emph{2016 {IEEE} {Power} and {Energy}
  {Society} {General} {Meeting} ({PESGM})}, Jul. 2016.

\bibitem{baran_optimal_1989}
\BIBentryALTinterwordspacing
M.~E. Baran and F.~F. Wu, ``Optimal sizing of capacitors placed on a radial
  distribution system,'' \emph{Power Delivery, IEEE Transactions on}, vol.~4,
  no.~1, pp. 735--743, 1989. [Online]. Available:
  \url{http://ieeexplore.ieee.org/xpls/abs_all.jsp?arnumber=19266}
\BIBentrySTDinterwordspacing

\bibitem{monticelli_observability_1986}
A.~Monticelli and F.~F. Wu, ``Observability {Analysis} for {Orthogonal}
  {Transformation} {Based} {State} {Estimation},'' \emph{IEEE Transactions on
  Power Systems}, vol.~1, no.~1, pp. 201--206, Feb. 1986.

\bibitem{noauthor_dataport_2017}
\BIBentryALTinterwordspacing
``Dataport,'' 2017. [Online]. Available: \url{http://www.pecanstreet.org/}
\BIBentrySTDinterwordspacing

\bibitem{noauthor_ieee_2017}
\BIBentryALTinterwordspacing
``{IEEE} {Distribution} {Test} {Feeders},'' 2017. [Online]. Available:
  \url{http://ewh.ieee.org/soc/pes/dsacom/testfeeders/}
\BIBentrySTDinterwordspacing

\bibitem{schneider_analytic_2018}
K.~P. Schneider, B.~A. Mather, B.~C. Pal, C.-W. Ten, G.~J. Shirek, H.~Zhu,
  J.~C. Fuller, J.~L.~R. Pereira, L.~F. Ochoa, and L.~R. De~Araujo, ``Analytic
  {Considerations} and {Design} {Basis} for the {IEEE} {Distribution} {Test}
  {Feeders},'' \emph{IEEE Transactions on Power Systems}, vol.~33, no.~3, pp.
  3181--3188, 2018.

\end{thebibliography}

\appendices

%










\end{document}